\documentclass[iop]{emulateapj}
\slugcomment{Accepted by  {\it The Astrophysical Journal}}
\newcommand{\angstrom}{\text{\normalfont\AA}}
\newcommand{\rc}{\text{$\chi^{2}$/$\nu${}}}

\def\la{\mathrel{\hbox{\rlap{\hbox{\lower4pt\hbox{$\sim$}}}\hbox{$<$}}}}
\def\ga{\mathrel{\hbox{\rlap{\hbox{\lower4pt\hbox{$\sim$}}}\hbox{$>$}}}}

\usepackage{gensymb}
\usepackage{amsmath}
\usepackage{amssymb}
\usepackage{graphicx}
\usepackage{wasysym}
\usepackage{url}

\shortauthors{Bhalerao}
\shorttitle{G292.0+1.8 X-ray Kinematics}

\newcommand{\kms}{~{km\,s$^{-1}$}}

\begin{document}

\title{X-Ray Ejecta Kinematics of the Galactic Core-Collapse Supernova Remnant G292.0+1.8}

\author{Jayant Bhalerao\altaffilmark{1}, Sangwook Park\altaffilmark{1}, Daniel Dewey\altaffilmark{2}, John P. Hughes \altaffilmark{3}, Koji Mori\altaffilmark{4}, and Jae-Joon Lee\altaffilmark{5}}

\altaffiltext{1}{Box 19059, Department of Physics, University of Texas at Arlington, Arlington, TX 76019; jayant.bhalerao@mavs.uta.edu}

\altaffiltext{2}{MIT Kavli Institute, Cambridge, MA 02139}
\altaffiltext{3}{Department of Physics and Astronomy, Rutgers University, 136 Frelinghuysen Road, Piscataway, NJ 08854-8019}

\altaffiltext{4}{Department of Applied Physics, University of Miyazaki, 1-1 Gakuen Kibana-dai Nishi, Miyazaki, 889-2192, Japan}
\altaffiltext{5}{Korea Astronomy and Space Science Institute, Daejeon, 305-348, Korea}

\begin{abstract}
We report on the results from the analysis of our 114 ks \textit{Chandra} HETGS observation of the Galactic core-collapse supernova remnant G292.0+1.8.  To probe the 3D structure of the clumpy X-ray emitting ejecta material in this remnant, we measured Doppler shifts in emission lines from metal-rich ejecta knots projected at different radial distances from the expansion center. We estimate radial velocities of ejecta knots in the range of -2300 $\lesssim$ $v_r$ $\lesssim$ 1400 km s$^{-1}$. The distribution of ejecta knots in velocity vs. projected-radius space suggests an expanding ejecta shell with a projected angular thickness of $\sim$90$\arcsec$ (corresponding to $\sim$3 pc at d = 6 kpc). Based on this geometrical distribution of the ejecta knots, we estimate the location of the reverse shock approximately at the distance of $\sim$4 pc from the center of the supernova remnant, putting it in close proximity to the outer boundary of the radio pulsar wind nebula. Based on our observed remnant dynamics and the standard explosion energy of $10^{51}$ erg, we estimate the total ejecta mass to be $\lesssim$ 8 M$_{\astrosun}$, and we propose an upper limit of $\lesssim$ 35 M$_{\astrosun}$ on the progenitor's mass.

\end{abstract}


\section {\label {sec:intro} INTRODUCTION}

G292.0+1.8 is a Galactic oxygen-rich (O-rich) core-collapse supernova remnant (CC SNR) that has been studied at different wavelengths over the past five decades. Previous studies have captured a complex portrait composed of typical elements for a CC SNR  \textendash {} a pulsar (Camilo et al. 2002; Hughes et al. 2003), and its wind-blown nebula or pulsar wind nebula (PWN, Hughes et al. 2001; Gaensler \& Wallace 2003 (GW03 hereafter); Park et al. 2007), the blast wave-shocked circumstellar medium (CSM, Park et al. 2002; Lee et al. 2009; Lee et al. 2010 (L10 hereafter)), and metal-rich ejecta knots strewn across the interior in intricate filamentary networks (Gonzalez \& Safi-Harb 2003; Park et al. 2004; 2007; Ghavamian et al. 2005; 2009; 2012; Winkler \& Long 2006; Winkler et al. 2009). Yet details about the progenitor star and how its explosion led to the complex patterns of shocked ejecta and CSM seen in the sky today, remain elusive. The mass of the progenitor star has not been tightly constrained ($\sim$20-40 M$_{\astrosun}$, Hughes \& Singh 1994; Gonzales \& Safi-Harb 2003; Park et al. 2004; L10; Kamitsukasa et al. 2014; Yang et al. 2014). It is unclear if the progenitor has gone through phases other than the red supergiant (RSG). The presence of the equatorial belt (a bright, belt-like emission feature of shocked dense CSM enhanced along the ``equator'' of the SNR, Park et al. 2002; Ghavamian et al. 2005; Lee et al. 2009) suggests that the progenitor was probably rapidly rotating and/or in a binary system, but extensive studies on the progenitor system have not been performed. The associated pulsar (PSR J1124-5916) is apparently off the geometric center of the SNR (e.g., Hughes et al. 2001) indicating a significant pulsar-kick which could be related to a non-symmetric SN explosion (Park et al. 2007). The details of the pulsar-kick and its relationships with the progenitor system and explosion mechanism in G292.0+1.8 are not known. In contrast to Cassiopeia A (Cas A, a $\sim$10 times younger cousin of G292.0+1.8), in which abundant Fe-group ejecta material is observed (e.g., Hwang \& Laming 2012, HL12 hereafter), such explosive nucleosynthesis products had not been detected in G292.0+1.8. Recently, a \textit{Suzaku} study detected faint Fe K-shell line emission in G292.0+1.8, probably originating from hot Fe-rich ejecta (Kamitsukasa et al. 2014).

A supernova (SN) explosion releases elements synthesized by the life-long efforts of a star (somewhat modified during its explosion) as metal-rich ejecta gas that expands into the surrounding CSM.  The interaction of the rapidly expanding ejecta with the surrounding CSM creates two powerful shock fronts: an outward-moving forward shock (FS) that heats the CSM, and an inward-moving reverse shock (RS) that propagates back heating the metal-rich ejecta near the SNR center (e.g., a recent review by Dewey 2010). The FS is clearly identified in G292.0+1.8 as the outermost boundary of the remnant in X-rays (L10), in radio (GW03), and in infrared (Lee et al. 2009; Ghavamian et al. 2009; 2012). The location of the RS is difficult to ascertain because the 3D ejecta distribution is projected on the plane of the sky. 

A useful method to probe the 3D structure of an SNR is to study the line-of-sight distribution of fast-moving ejecta knots by measuring their radial velocities ($v_r$). Mapping the 3D distribution of ejecta may help locate the RS front. The High Energy Transmision Grating Spectrometer (HETGS) on board \textit{Chandra} provides a powerful high resolution spectroscopy to estimate Doppler shifts in the X-ray spectral lines of metal-rich ejecta knots, a measure of their $v_r$. The utility of this method has been successfully demonstrated with the bright ejecta-dominated SNR Cas A (e.g., Lazendic et al. 2006). Based on our \textit{Chandra} HETGS observations, we apply a similar method to map the $v_r$ distribution of 33 bright knots and filaments in G292.0+1.8. Here we provide the first insight into the 3D internal architecture of this textbook-type CC SNR in X-rays.

\section{\label{sec:obs} OBSERVATIONS \& DATA REDUCTION}

We performed our \textit{Chandra} HETGS observation of G292.0+1.8 between 2011 March 20 and 2011 March 27. The aim point was set at RA(J2000.0) = 11$^h$ 24$^m$ 39$^s$.5, Dec(J2000.0) = -59\degree 15$\arcmin$ 56$\arcsec$.40 to detect a majority of bright ejecta knots within $\sim$2$\arcmin$ off-axis. The observation was composed of three ObsIDs (12555, 13242, and 13243). In each observation all six ACIS-S CCDs were operated in the full-frame readout mode. We processed the raw event files using  CIAO version 4.4 and CALDB version 4.4.3. We followed the standard data reduction methods involving grade and hot pixel filtering. We found no significant contamination from flaring background. We processed each ObsID individually, and all three ObsIDs were combined for data analysis to yield a total effective exposure of $\sim$114 ks. As supplementary data (see \S{} 3), we also used the archival ACIS-I data of G292.0+1.8 (Park et al. 2007). We reprocessed all six ObsIDs of the ACIS-I data following standard data reduction procedures with CIAO version 4.3 and CALDB version 4.4.3, which resulted in a total effective exposure of 509 ks.

\section{\label{sec:result} ANALYSIS \& RESULTS}

We extracted source spectra from numerous small regions in G292.0+1.8, and measured line center energies using methods similar to those applied for the study of ejecta knots in Cas A (Lazendic et al. 2006; Rutherford et al. 2013). We used a fixed order-sorting range of $\pm$10\% to extract the first-order spectrum (Figure 1). We created the zeroth-order image of the ObsID with the longest exposure (ObsID 12555) in the 0.8\textendash{}2.2 keV band in which bright K$\alpha$ lines from He- and H-like Ne, Mg and Si ions are present. Based on this image we identified the zeroth-order locations of bright, compact knots which would have small cross-dispersion widths ($\sim$2$\arcsec$ \textendash{} 9$\arcsec$ in angular sizes, and $\sim$4$\arcsec$ on average, for which the angular dispersion of these small source regions do not affect our Doppler line shift measurements). Using these line centers and cross-dispersion widths, we extracted the dispersed spectra from these small knots from all three ObsIDs applying standard CIAO tools \textendash{}~TGCat scripts.\footnote{\url{http://tgcat.mit.edu/}} We show an example of an HETGS spectrum extracted from a small bright emission feature in Figure 1.

We analyzed the first-order spectra, corresponding to orders MEG~$\pm$1 and HEG~$\pm$1, using custom scripts executed in the ISIS software package\footnote{\url{http://space.mit.edu/CXC/ISIS/}} (Houck \& Denicola 2000). For each knot, we combined the spectra extracted from all three ObsIDs. Five emission lines are useful for Doppler shift measurements of individual ejecta knots: atomic emission lines from the K-shell transitions in the He- and H-like ions of Ne and Mg, and in the He-like Si. The rest wavelengths for these lines are listed in Table 1. We detect and characterize these lines in the dispersed spectra of small knots in G292.0+1.8 using simple phenomenological model fits applied to a narrowband around each line. Our model consists of two Gaussians for the $\mathrm{Ly\alpha}$ lines (one for the line and the other to approximate the underlying continuum) and four Gaussians for the $\mathrm{He\alpha}$ triplets (three corresponding to the forbidden (\textit{f}), intercombination (\textit{i}) and resonance (\textit{r}) lines, and one for the underlying continuum). We use a sum of broad Gaussians to approximate the underlying continua in the five line-regions that we fit: the Gaussians provide a computationally simple method that allows each local continuum level to be adjusted with reasonable independence, since the Gaussians decrease quickly outside of their wavelength ranges. Free parameters in our model are the line center, the line flux, the line width ($\sigma$), and the continuum flux. For the continuum Gaussian component, we fixed the center energy at the rest wavelength of the corresponding line while varying the area of the Gaussian. The model fits for the $\mathrm{He\alpha}$ triplets have the same degrees of freedom as those of the $\mathrm{Ly\alpha}$ lines, because the wavelength and fluxes of the \textit{f} and \textit{i} lines are set to be proportional to those of the \textit{r} line. The \textit{i/r} and \textit{f/r} flux ratios we used were based on the observed values for Capella and SN 1987A (Canizares et al. 2000; Dewey et al. 2008) in which we assumed a low-density gas (which should also be the case for G292.0+1.8). We list these flux ratios in Table 1. We note that our primary goal of line shift measurements is not very sensitive to the exact ratios between these triplet lines. Also, the counting statistics dominate the observed line fluxes, and our Doppler velocity shift measurements are based on several emission line complexes (Table 2). For our Doppler velocity shift measurements, we first fitted each of the five lines listed above to detect a valid line feature. We scaled all the lines in the model to the model wavelength of the Ne IX line. The detected lines were then jointly fitted to estimate a common velocity shift. We fitted 65 small knots with these models, and used 33 knots that show statistically acceptable fits (\rc\ $\textless$ 2, for a combined fit of all detected lines) for our Doppler shift measurements. We excluded 32 knots from our $v_r$ measurements because their low signal-to-noise ratio did not allow us to detect valid line features. We show extracted spectra and best-fit models for three example regions in Figure 2. Based on the shifts in our measured line centers from the rest wavelengths, we estimate $v_r$ for these knots (Table 2).

To identify the origins (shocked ejecta vs.\ CSM) of these 33 knots, we investigated their spectral properties using our deep 509 ks ACIS-I observation of G292.0+1.8 (Park et al. 2007). We used the ACIS data to utilize the significantly higher photon statistics (by more than an order of magnitude in the 0.3-5 keV band) than those in the HETG data . We performed spectral model fits for the observed ACIS spectra of these 33 regions to measure their metal abundances. For these spectral model fits we subtracted the background emission spectrum using spectra extracted from nearby source-free (dark, \textit{ejecta-free}) regions within the SNR. We performed spectral model fits using the absorbed (\textit{phabs} in XSPEC) non-equilibrium ionization (NEI) plane-parallel shock model (Borkowski et al. 2001) with variable abundances (\textit{vpshock}, NEI version 2.0 with augmented ATOMDB, Smith et al. 2001; Badenes et al. 2006). We added a power law component for regions projected within or near the PWN. We varied O, Ne, Mg, Si, S and Fe abundances while fixing other elemental abundances at solar values (Anders \& Grevesse 1989). Based on these abundance measurements we identified 24 ejecta knots (showing abundances typically $>$ several times solar for one or more elements). We identified 9 CSM-like features with sub-solar abundances for all fitted elements (Table 2). Most of the CSM features are positioned along the equatorial belt. 

We constructed a $v_r$\textendash $r_p$ distribution for these 33 knots (Figure 3), where $r_p$ is the projected distance from the expansion center. For a homologous expansion of ejecta knots in G292.0+1.8, the 3D spatial velocities ($v_{3D}$) of individual ejecta knots are proportional to their physical distances or 3D radii ($r_{3D}$) from the expansion center. The constant relating this proportionality is $r_{3D}/v_{3D}$ = 0.1055\,$\mathrm{\arcsec/km\,s^{-1}}$, assuming an expansion age of 3000 yr (Winkler et al. 2009) and a distance to the SNR of 6 kpc (GW03). Knots at the same $r_{3D}$ will differ in their $r_p$ and $v_r$ values depending on their projected locations. In Figure 3 we overlay four elliptical loci to relate $v_r$ and $r_p$ from the SNR's expansion center, assuming this proportionality constant. The smallest elliptical locus corresponds to a physical distance ($r_{3D}$ from the expansion center) of $\sim$3.5 pc (at the projected angular distance $\sim$120\arcsec) and roughly represents the angular size of the radio PWN, GW03). The next two loci at $\sim$3.8 pc (at $\sim$130\arcsec) and $\sim$6.4 pc (at $\sim$220\arcsec) have been qualitatively estimated by eyeball inspection to contain the majority of the ejecta knots within a shell. The outermost locus at $\sim$7.7 pc ($\sim$265\arcsec\,) corresponds to the FS (L10). We roughly estimate (by eyes) the velocity centroid at +150 km s$^{-1}$ which is similar to that estimated in the optical band (Ghavamian et al. 2005). We show the projected positions for the 33 regional features and these elliptical loci in Figure 4.

\section{\label{sec:disc} DISCUSSION}

Our estimated radial velocity range of -2300 $\lesssim$  $v_r$ $\lesssim$ 1400 km s$^{-1}$ for X-ray ejecta knots is in plausible agreement with earlier optical measurements of ejecta velocities in G292.0+1.8. Ghavamian et al. (2005) reported ejecta radial velocities in the range of -1700 $\lesssim$ $v_r$ $\lesssim$ +1700 km s$^{-1}$ for O-rich optical ejecta knots in G292.0+1.8. Winkler et al. (2009) conducted proper motion studies of O-rich knots in the optical band, and measured east-west velocities of -1800 d$_6$ \textless {} $v_x$ \textless {} 1490 d$_6$ km s$^{-1}$ and north-south velocities in the range of -3570 d$_6$ \textless {} $v_y$ \textless {} 2340 d$_6$ km s$^{-1}$, where d$_6$ is the distance to G292.0+1.8 in units of 6 kpc. While X-ray and optical emissions originate from ejecta gas with different thermal conditions, and thus X-ray ejecta knots generally do not show optical counterparts, we detect some spatial correlations between X-ray and optical ejecta knots. The highly redshifted knot E7 shows positional coincidence with the largely redshifted optical \textquotedblleft spur\textquotedblright\, in the southeast region of the SNR, and the blueshifted knots E5 and E11 are in similar positions to blueshifted optical knots in the northern parts of the SNR. Spatial correlation between X-ray and optical emission is also supported by the observation that several X-ray filamentary structures in the north coincide with optical knots located near their termini (Figure 13 in Ghavamian et al. 2005). Thus, in G292.0+1.8 the ejecta gas at various thermal states appears to share some bulk motion.

We detect a significantly larger number of blueshifted knots than redshifted ones (17 of 24 ejecta knots are blueshifted). For the blueshifted ejecta knots, we also estimate generally higher velocity magnitudes than the redshifted ones: e.g., seven blueshifted knots show $v_r$ \textgreater {} 1000 km s$^{-1}$ while only one redshifted ejecta knot shows such a high $v_r$. A similar non-symmetric $v_r$ distribution of ejecta in G292.0+1.8 was observed in the optical band, where a significantly larger number of blueshifted knots was detected, especially in the north (Ghavamian et al. 2005). Asymmetries in $v_r$ have also been seen in other O-rich SNRs, for which interpretations included asymmetric SN explosions and density variations in the CSM. SNR 1E 0102.2\textendash 7219, in the Small Magellanic Cloud, shows a larger number of blueshifted bright knots but the redshifted knots show generally higher $v_r$ (Vogt \& Dopita 2010). SNR 0540-69.3, in the Large Magellanic Cloud (LMC), shows a generally redshifted spectrum of ejecta (Kirshner et al. 1989). SNR N132D, also in the LMC, shows higher $v_r$ in its blueshifted ejecta (Vogt \& Dopita 2011). Cas A shows higher $v_r$ in its redshifted ejecta (Milisavljevic \& Fesen 2013 and references therein).

Possible origins for the observed $v_r$ asymmetry in G292.0+1.8 may include several scenarios such as an asymmetric SN explosion, CSM density variations (near vs. far sides of the SNR) along the line-of-sight, a clumpiness variation of the ejecta, and self-absorption of redshifted emission.  We discuss each of these scenarios below.  An asymmetric SN explosion may have channeled more kinetic energy towards the Earth along the line-of-sight. Observational evidence supporting an asymmetric SN explosion for G292.0+1.8 has been reported in previous works: e.g., higher X-ray ejecta temperatures in the northwest than in the southeast regions (Park et al. 2007),  the absence of Si emission in the southeast (Park et al. 2002; Ghavamian et al. 2012), higher proper motions of optical ejecta knots along the north-south than in the east-west directions (Winkler et al. 2009), and the $\gtrsim$1 pc displacement (to southeast from the SNR's expansion center) of the associated pulsar PSR J1124-5916 (e.g., Winkler et al. 2009). In such an asymmetric SN explosion, the energy output might have resulted in a larger amount of blueshifted fast-moving ejecta material as observed in X-rays (this work) and in optical (Ghavamian et al. 2005).

Another tentative scenario for the observed $v_r$ asymmetry could be a non-uniform CSM. For instance, a significant CSM density variation between the near and far sides of the SNR might have created asymmetry in the RS structure, causing a greater inward migration of the RS on the near side (if the CSM density is higher there) than the far side, thus interacting with more ejecta material to produce more blueshifted material. A CSM density variation in G292.0+1.8 is suggested by large filamentary structures such as the equatorial belt (Park et al. 2004; Lee et al. 2009; L10; Ghavamian et al. 2012), and by a non-uniform circumstellar environment as seen in the mid-infrared (Park et al. 2007). Some azimuthal CSM density variation has been observed in G292.0+1.8 with regions in the southeast showing lower CSM densities than other regions (L10). However, it is not clear if this azimuthal CSM density structure orginated from variation in the progenitor's wind density or from an asymmetric SN explosion. Also, a deeper migration of the RS on the near side of the SNR would create more blueshifted material with lower $v_r$ from the heating of slower moving central ejecta regions. One would therefore expect to see a larger number of low $v_r$ blueshifted ejecta regions projected near the SNR center, which is not clearly evident. Hence, the presence and contribution of a CSM density variation along the line-of-sight between the near and far sides of the SNR that would result in the observed blueshift-dominated ejecta in G292.0+1.8 is unclear, although it cannot be ruled out.

A selection effect due to a clumpiness variation of the metal-rich ejecta in the SNR might also have contributed to the observed blueshift predominance in G292.0+1.8. Since we are more likely to select small bright knots (for our $v_r$ measurements) that would originate in clumpier regions than in smoother plasma, an SNR with a substantially larger number of clumpy ejecta features on the near side could result in the observed blueshift predominance. An asymmetric clumpy ejecta distribution has been proposed to explain observed optical emission line asymmetries in SNe 1993J (Spyromilio 1994) and 1990I (Elmhamdi et al. 2004), with further support in theoretical studies (e.g., Herrington et al. 2010).

The observed $v_r$ asymmetry in G292.0+1.8 might have originated from self-absorption of redshifted emission by material within the SNR (e.g., ejecta dust, Ghavamian et al. 2009; 2012). However, there is no observational evidence for significant self-absorption of X-ray emission that could lead to the observed $v_r$ asymmetry in G292.0+1.8: e.g., we find that the column density ($N_{H}$) for the highly redshifted knot E7 is consistent with that for blueshifted regions. In general the existing observational evidence appears to favor an asymmetric explosion scenario for the observed $v_r$ asymmetry. However, the true origin remains elusive.

Most of the ejecta knots in our study occupy a thick shell in $v_r$\textendash $r_p$ space, corresponding to the RS-heated hot ejecta gas (solid diamonds in Figure 3). On the other hand, most of the CSM filaments occupy the low $v_r$ region (open circles in Figure 3), and are positioned along the equatorial belt (Figure 4). Unless our 24 ejecta knots represent a heavily biased sample, the inner and outer radii of this ejecta shell may roughly correspond to the locations of the RS ($r_{in}$ $\sim$ 130$\arcsec$) and contact discontinuity (CD, $r_{out}$ $\sim$ 220$\arcsec$) respectively (the projected angular distances $r_{in}$ and $r_{out}$ are measured from the explosion site determined from the proper motions of optical ejecta knots (Winkler et al. 2009)). We find that nearly all of 62 fast-moving optical ejecta knots (Ghavamian et al. 2005) also lie within our X-ray-estimated ejecta shell (we estimate that only two of them are positioned at a slightly larger radius than our CD), further supporting our inferred location of the RS. We note that there are a few regions with large uncertainties in $v_r$, likely due to the relatively weak emission lines in these features. For example, $v_r$ measurements for regions C28, E29, C31 and E33 were based on only one, relatively faint line ($\mathrm{Si\:He\alpha}$, Table 2).

Our estimate of the RS location gives a ratio between the radii of the RS and FS, R$_\text{RS}$/R$_\text{FS}$ $\sim$130$\arcsec$/265$\arcsec$ $\sim$ 0.5. This ratio is consistent with previous estimates at other wavelengths: $\sim$0.47 by Braun et al. (1986) based on radio and infrared data, and $\sim$0.5 by GW03, based on radio data. This R$_\text{RS}$/R$_\text{FS}$ ratio in G292.0+1.8 is similar to that seen in other young O-rich SNRs: e.g., $\sim$0.6-0.8 for Cas A (at age $\sim$ 330 yr, HL12) and $\sim$0.5-0.7 for 1E 0102.2\textendash 7219 (at age $\sim$ 1000 yr, Gaetz et al. 2000; Flanagan et al. 2004). A smaller ratio  of $\sim$0.3 was estimated in N132D (age $\sim$ 2500 yr) suggesting a dynamically more evolved stage for this SNR, with its RS possibly accelerating towards the SNR center (Vogt \& Dopita 2011). Our estimated R$_\text{RS}$/R$_\text{FS}$ ratio for G292.0+1.8 is significantly smaller than the values predicted by self-similar solutions (R$_\text{RS}$/R$_\text{FS}$ $>$ 0.7, Chevalier 1982), suggesting that this SNR has evolved beyond the early ejecta-dominated phase. Truelove \& McKee (1999) (TM99 hereafter) developed a hydrodynamic framework that extends the model to later times when the RS reaches the ejecta core region and the FS approaches the late-time Sedov-Taylor phase. They presented explicit results for the SNR evolution in a uniform density medium. Laming \& Hwang (2003) and HL12 (LH03-HL12 hereafter) extended the TM99 model for SNRs expanding into stellar winds with radial mass density profile $ \rho \propto r^{-2}$. Since G292.0+1.8 is expanding into an RSG wind (L10), we applied the LH03-HL12 model for G292.0+1.8 and successfully reproduce our estimated R$_\text{RS}$/R$_\text{FS}$  $\sim$ 0.5 at age = 3000 yr. The age of $\sim$3000 yr has been estimated for G292.0+1.8 based on kinematic studies (Ghavamian et al. 2005; Winkler et al. 2009), which is similar to the characteristic spin-down age of PSR J1124-5916 in G292.0+1.8 (2900 yr, Camilo et al. 2002). For these model calculations we assumed a canonical explosion energy E$_\text{0}$ = $1 \times 10^{51}$ erg, and a preshock CSM density 0.1-0.3 $\mathrm{ cm^{-3}}$ at R$_\text{FS}$ = 7.7 pc (L10). Based on the LH03-HL12 model, assuming the power-law index of 5 $\lesssim$ n $\lesssim$ 14 for the ejecta radial density profile in the outer layers (with an inner \textit{constant} density core) and SNR age of 2890 $\lesssim$ t $\lesssim$ 3080 yr (Winkler et al. 2009), we calculate the total ejecta mass $\mathrm{M_{ej}}$ $\lesssim$ 8 M$_{\astrosun}$. Combining our upper limit for $M_{ej}$ with our previously determined averaged wind mass estimate of $M_w$ $\sim$ 25 M$_{\astrosun}$ (L10), we suggest an upper limit of $\sim$35 M$_{\astrosun}$ for the progenitor mass ($M_{prog}$) of G292.0+1.8. This upper limit provides a constraint on $M_{prog}$ based on the observed dynamics of the SNR, and is in plausible agreement with previous nucleosynthesis-based estimates for $M_{prog}$: e.g., 25 $M_{\astrosun}$ (Hughes \& Singh 1994; Park et al. 2004), 30-40 $M_{\astrosun}$ (Gonzales \& Safi-Harb 2003), 30-35 $M_{\astrosun}$ (Kamitsukasa et al. 2014) and 25-30 $M_{\astrosun}$ (Yang et al. 2014).

Considering the sharp boundary between the PWN and the outer plateau in radio, GW03 suggested an RS-PWN interaction, probably in the very early stage where the RS has not compressed the PWN significantly. On the other hand, a large pressure difference between the PWN and the thermal gas in X-rays had suggested that the RS and PWN had not yet interacted (Park et al. 2004). Also, a large O/Ne mass ratio in the mid-infrared suggested that the inner explosive nucleosynthesis products might not have undergone significant mixing, and could still remain unshocked (Ghavamian et al. 2009), which generally supports a non-interaction between the RS and PWN. Our estimated RS location is overall close to the outer boundary of the radio PWN (Figure 4). The position of the X-ray PWN in G292.0+1.8 is generally consistent with its radio counterpart, and the projected angular extent of the X-ray PWN is smaller than that of the radio PWN (Figure 5). This X-ray-radio PWN size difference is consistent with the standard picture of more effective synchrotron loss of X-ray emission in the outer layers of PWNs (e.g., Gaensler \& Slane 2006). Theoretical and observational studies suggest that late stages of PWN-RS interactions are characterized by irregular PWN morphologies. These studies also suggest that asymmetry in the RS structure of SNRs will result in displacement of PWNs relative to their PSRs, and inconsistencies in the sizes and positions of radio PWNs and their X-ray counterparts in late PWN-RS interaction stages (e.g., Gaensler \& Slane 2006 and references therein). These signs of late-stage PWN-RS interactions are not clearly evident in G292.0+1.8. Therefore, a PWN-RS interaction, if it has started, should be in an early stage in this SNR (as suggested by GW03). We note that Park et al. (2004) estimated the thermal pressure of the SNR using only a small region on the equatorial belt. The RS front in G292.0+1.8 may not be smooth or spherically symmetric after interacting with a non-uniform CSM. Then we consider that the large pressure difference between the PWN and thermal gas estimated by Park et al. (2004) might have been a local effect. X-ray thermal pressure measurements from extensive areas of the SNR  would be helpful to test this discrepancy (which is beyond the scope of this paper, and will be included in our follow-up work on the ACIS-I data analysis). If the RS front is close to the PWN (and probably interacting with it), it may be nearing the SNR's central region where the Fe-rich ejecta might be expanding. A recent \textit{Suzaku} detection of the Fe K-shell emission line in the hot ejecta gas of G292.0+1.8 (Kamitsukasa et al. 2014) may support this scenario, opening further avenues in the quest to decipher this complex remnant.

\section{\label{sec:summ} SUMMARY}

Based on our $\sim$114 ks \textit{Chandra} HETGS observation of G292.0+1.8, we measure $v_r$ from Doppler line shifts for 33 bright knots in the SNR. Our measured $v_r$ is in the range of -2300 $\lesssim$  $v_r$ $\lesssim$ 1400 km s$^{-1}$. We detect a $v_r$ asymmetry with a larger number of blueshifted ejecta knots than redshifted ones. Our measured $v_r$ range and observed blueshifted ejecta knot predominance are generally consistent with results from optical observations (Ghavamian et al. 2005). We find that the blueshifted X-ray ejecta knots generally show higher velocity magnitudes than the redshifted ones. Other O-rich SNRs have also been found to show $v_r$ asymmetry. The cause for the $v_r$ asymmetry in G292.0+1.8 may have been an asymmetric SN explosion, although environmental effects such as CSM density variations along the line-of-sight cannot be ruled out. Based on the distribution of the ejecta knots in $v_r$\textendash $r_p$ space, we qualitatively locate the positions of the RS and CD. Our inferred RS position agrees with previous estimates based on radio and IR data, and with the hydrodynamic model for an SNR expanding into an RSG wind. Employing the SNR's dynamics we calculate the total ejecta mass of $\lesssim$ 8 M$_{\astrosun}$, and propose an upper limit of $\sim$35 M$_{\astrosun}$ for the G292.0+1.8 progenitor mass. Our inferred location of the RS places it in close proximity to the outer boundary of the PWN, suggesting the possibility of early-stage PWN-RS interactions, and the possible onset of inner Fe-rich ejecta-heating by the RS.

\acknowledgments

This work was supported in part by the SAO through \textit{Chandra} grant GO1-12077X.  JB acknowledges support from the NASA Texas Space Grant Consortium. DD was supported by NASA through SAO contract SV3-73016 to MIT for support of the \textit{Chandra} X-ray Center and Science Instruments.

\clearpage

\begin{deluxetable*}{cccc}
\tablewidth{0pt}
\tabletypesize{\scriptsize}
\tablecaption{Emission Lines used for Doppler Shift Measurements\label{tab:lines}}
\tablehead{
\colhead{Ion/transition} & 
\colhead{Rest Wavelength\tablenotemark{\,a} (\angstrom)} &
\multicolumn{2}{c}{Flux ratios} \\
\colhead{~~~} & 
\colhead{~~~} & 
\colhead{\textit{i}/\textit{r}\tablenotemark{\,b}}   &
\colhead{\textit{f}/\textit{r}\tablenotemark{\,b}} 
}
\startdata
 $\mathrm{Ne\: IX\: \textit{r}\phm{a} (Ne\: He\alpha\: \textit{r})\phm{a}}$ & 13.447 & 0.26 & 0.59 \\ 
 $\mathrm{Ne\: IX\: \textit{i}\phm{a} (Ne\: He\alpha\: \textit{i})\phm{a}}$ & 13.553 & \textendash & \textendash \\ 
 $\mathrm{Ne\: IX\: \textit{f}\phm{a} (Ne\: He\alpha\: \textit{f})\phm{a}}$ & 13.699 & \textendash & \textendash \\ 
 $\mathrm{Ne\: X\phm{a} (Ne\: Ly\alpha)~~~~~\phm{a}}$ & 12.134 & \textendash & \textendash \\ 
 $\mathrm{Mg\: XI\: \textit{r}\phm{a} (Mg\: He\alpha\: \textit{r})}$ & 9.169 & 0.21 & 0.43\\ 
 $\mathrm{Mg\: XI\: \textit{i}\phm{a} (Mg\: He\alpha\: \textit{i})}$ & 9.231 & \textendash & \textendash\\
 $\mathrm{Mg\: XI\: \textit{f}\phm{a} (Mg\: He\alpha\: \textit{f})}$ & 9.314 & \textendash & \textendash\\  
 $\mathrm{Mg\: XII\: (Mg\: Ly\alpha)~~~}$ & 8.421 & \textendash & \textendash \\ 
 $\mathrm{Si\: XIII\: \textit{r}\phm{a} (Si\: He\alpha\: \textit{r})}$ & 6.648 & 0.23 & 0.43\\
 $\mathrm{Si\: XIII\: \textit{i}\phm{a} (Si\: He\alpha\: \textit{i})}$ & 6.688 & \textendash & \textendash\\
 $\mathrm{Si\: XIII\: \textit{f}\phm{a} (Si\: He\alpha\: \textit{f})}$ & 6.740 & \textendash & \textendash\\  
\enddata
\tablenotetext{a}{The H-like $\mathrm{Ly\alpha}$ line values are from Johnson \& Soff 1985, and the He-like line values are from Drake 1988.} 
\tablenotetext{b}{The Ne IX and Si XIII ratios are based on the observed flux ratios for SN 1987A, and the Mg XI ratios are based on the observed flux ratios for Capella and SN 1987A (Canizares et al. 2000; Dewey et al. 2008). We assume a low-density gas which should also be the case for G292.0+1.8. The letters \textit{r}, \textit{i} and \textit{f} indicate the resonance, intercombination and forbidden transitions respectively in the He-like ions.}  
\end{deluxetable*}

\begin{deluxetable*}{cccrlccc}
\tablewidth{0pt}
\tabletypesize{\scriptsize}
\tablecaption{Radial Velocities of X-ray Emission Features in G292.0+1.8 \label{tab:knots}}
\tablehead{
\colhead{Knot} & 
\colhead{Arc seconds}  &
\colhead{Position}   &
\colhead{~~~~~$v_r$~} &
\colhead{$\pm$\ 90\%}   &
\colhead{$\chi^2/\nu$} &
\colhead{Lines} &
\colhead{Knot}  \\
\colhead{ID} & 
\colhead{(from center)\tablenotemark{\,a}}  &
\colhead{angle (deg)\tablenotemark{\,b}}   &
\multicolumn{2}{c}{---~~\kms~~---} &
\colhead{} &
\colhead{used\tablenotemark{\,c}} &
\colhead{origin} 
}
\startdata

 E\,1 & 125.8 & 102.4 & 41 & $\pm$\,297 & 1.59 &  1,\,2,\,3,\,4  & Ejecta  \\ 
 E\,2 & 182.7 & 80.9 & -496 & $\pm$\,289 & 1.87 &  1,\,2,\,3,\,4  & Ejecta  \\ 
 C\,3 & 221.9 & 252.1 & 677 & $\pm$\,751 & 1.30 &  2,\,3,\,5  & CSM  \\ 
 E\,4 & 96.3 & 181.3 & -2221 & $\pm$\,407 & 1.48 &  2,\,3,\,4  & Ejecta  \\ 
 E\,5 & 134.0 & 48.9 & -981 & $\pm$\,435 & 1.60 &  2,\,3,\,4  & Ejecta  \\ 
 C\,6 & 62.5 & 90.0 & -503 & $\pm$\,2810 & 1.34 &  1,\,2  & CSM  \\ 
 E\,7 & 54.3 & 117.7 & 1396 & $\pm$\,271 & 1.68 &  1,\,2,\,3,\,4  & Ejecta  \\ 
 C\,8 & 15.4 & 97.5 & -792 & $\pm$\,655 & 1.94 &  2,\,3,\,5  & CSM  \\ 
 C\,9 & 43.1 & 276.7 & -990 & $\pm$\,693 & 1.29 &  1,\,2,\,3,\,5  & CSM  \\ 
 E\,10 & 92.1 & 248.9 & -1822 & $\pm$\,719 & 1.30 &  1,\,2,\,3,\,5  & Ejecta  \\ 
 E\,11 & 75.0 & 330.5 & -1289 & $\pm$\,923 & 1.39 &  2,\,5  & Ejecta  \\ 
 E\,12 & 69.7 & 306.2 & -1623 & $\pm$\,342 & 1.40 &  1,\,3,\,5  & Ejecta  \\ 
 E\,13 & 156.0 & 324.8 & -479 & $\pm$\,235 & 1.40 &  1,\,2,\,3,\,4,\,5  & Ejecta  \\ 
 E\,14 & 198.2 & 307.3 & -269 & $\pm$\,633 & 1.88 &  2,\,3,\,4  & Ejecta  \\ 
 E\,15 & 172.1 & 301.2 & -78 & $\pm$\,332 & 1.59 &  1,\,2,\,3,\,4,\,5  & Ejecta  \\ 
 C\,16 & 56.8 & 269.0 & 107 & $\pm$\,410 & 1.51 &  1,\,2,\,3,\,5  & CSM  \\ 
 E\,17 & 203.2 & 333.2 & 320 & $\pm$\,413 & 1.57 &  2,\,3,\,4,\,5  & Ejecta  \\ 
 E\,18 & 193.4 & 24.0 & -115 & $\pm$\,603 & 1.15 &  3,\,5  & Ejecta  \\ 
 E\,19 & 57.3 & 210.8 & -453 & $\pm$\,712 & 1.57 &  2,\,3,\,4  & Ejecta  \\ 
 E\,20 & 209.9 & 9.0 & 785 & $\pm$\,1048 & 1.01 &  1,\,2  & Ejecta  \\ 
 E\,21 & 147.3 & 289.9 & 381 & $\pm$\,657 & 1.66 &  1,\,2,\,3,\,4  & Ejecta  \\ 
 E\,22 & 203.0 & 355.8 & 839 & $\pm$\,452 & 1.33 &  2,\,3  & Ejecta  \\ 
 E\,23 & 140.9 & 143.0 & -1346 & $\pm$\,313 & 1.05 &  2,\,3,\,4  & Ejecta  \\ 
 E\,24 & 145.1 & 144.0 & -1074 & $\pm$\,271 & 1.14 &  1,\,2,\,3,\,4  & Ejecta  \\ 
 E\,25 & 56.5 & 295.2 & -1007 & $\pm$\,528 & 1.85 &  2  & Ejecta  \\ 
 E\,26 & 147.6 & 322.7 & -610 & $\pm$\,246 & 1.29 &  1,\,2,\,3,\,4  & Ejecta  \\ 
 E\,27 & 163.1 & 328.9 & -230 & $\pm$\,389 & 1.19 &  2,\,3,\,4  & Ejecta  \\ 
 C\,28 & 123.3 & 264.4 & -3087 & $\pm$\,1502 & 1.26 &  5  & CSM  \\ 
 E\,29 & 172.2 & 245.2 & 394 & $\pm$\,1763 & 1.15 &  5  & Ejecta  \\ 
 C\,30 & 85.6 & 265.3 & -420 & $\pm$\,778 & 1.04 &  2  & CSM  \\ 
 C\,31 & 109.6 & 52.2 & -811 & $\pm$\,1528 & 1.74 &  5  & CSM  \\ 
 C\,32 & 129.6 & 353.5 & -405 & $\pm$\,765 & 1.57 &  1,\,3  & CSM  \\ 
 E\,33 & 76.9 & 336.0 & -2301 & $\pm$\,1799 & 1.39 &  5  & Ejecta  \\

\enddata
\tablenotetext{a}{Angular distance from the  
the optical expansion center given by Winkler et al. 2009: 
R.A.\ $=$ 11$^h$24$^m$34.4$^s$, Decl.\ $=$ $-$59$^\circ$15'51'' (J2000);}
\tablenotetext{b}{Measured counterclockwise, north to east;}
\tablenotetext{c}{The lines used in fitting are:
1 = Ne IX, 2 = Ne X, 
3 = Mg XI, 4 = Mg XII and 5 = Si XIII.}
%
\end{deluxetable*}


\begin{figure*}[]
\figurenum{1}
\begin{center}
\includegraphics[angle=0,scale=0.55]{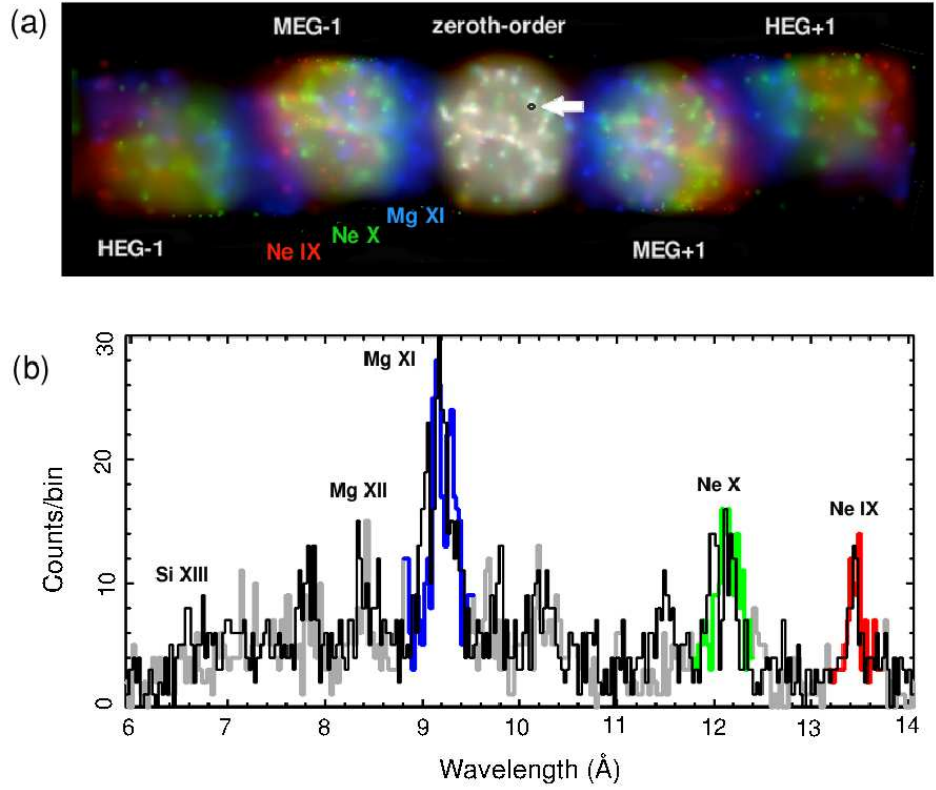}\\
\end{center}
\figcaption[]{{(\textit{a}): Dispersed image of the spectrum for G292.0+1.8 showing the zeroth order in the center and the first order, color coded by energy. Red: Ne IX (0.90\textendash{}0.93 keV), green: Ne X (1.02\textendash{}1.06 keV) and blue: Mg XI (1.33\textendash{}1.38 keV). (\textit{b}): Combined MEG spectrum for knot E13 (identified by an arrow on the zeroth order image in (\textit{a})), showing the Si XIII, Mg XII, Mg XI, Ne X and Ne IX lines used in the Gaussian fit. Black: MEG -1, gray: MEG +1. For comparisons, the MEG +1 data (gray) corresponding to the Mg XI, Ne X and Ne IX lines in the lower panel are highlighted using the same color scheme as in the top panel.}\label{fig:Figure 1}}
\end{figure*}

\begin{figure*}[]
\figurenum{2}
\begin{center}

\includegraphics[angle=0,scale=0.21]{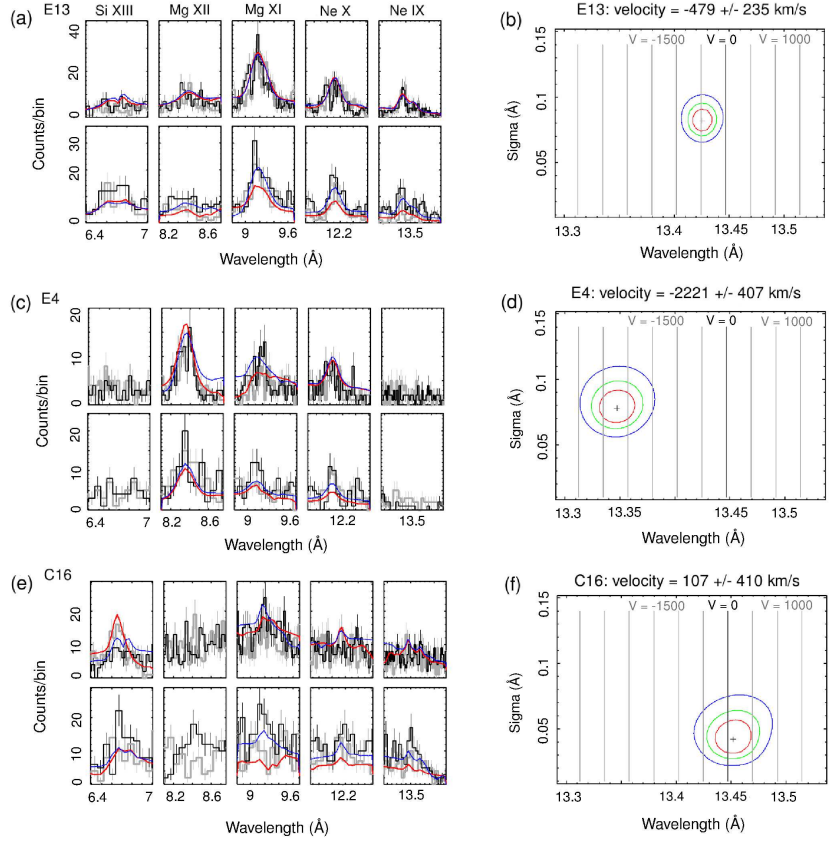}

\end{center}

\figcaption[]{{\textit{Left} (\textit{a, c, e}): Best-fit joint Gaussian model fits for detected lines (Si XIII, Mg XII, Mg XI, Ne X and Ne IX) for three sample regions. For each region MEG spectra are in the upper panel and HEG spectra are in the lower panel. Black: -1 order data, gray: +1 order data, blue: best-fit model for -1 order, red: best-fit model for +1 order. \textit{Right} (\textit{b, d, f}): confidence contour plots (68\% (red), 90\% (green), 99\% (blue)) for the  combined fitting of all detected lines. To make these contour plots we first fitted each of the five lines to detect a valid line feature. The detected lines (shown with red and blue model fit curves on the left), were then fitted jointly to estimate a common velocity shift. All of the lines in the model are scaled to the wavelength of the Ne IX model line. The wavelength of the Ne IX line center is plotted on the horizontal axis, and the common width of the lines is on the vertical axis. Starting at the top the regions are (\textit{a, b}): knot E13, a region with a large radial distance and low velocity, (\textit{c, d}): knot E4, a region with a small radial distance and high velocity, and (\textit{e, f}): knot C16, a CSM filament located at the equatorial belt showing a low velocity.}\label{fig:Figure 2}}
\end{figure*}

\begin{figure*}[]
\figurenum{3}
\begin{center}
\includegraphics[angle=0,scale=0.7]{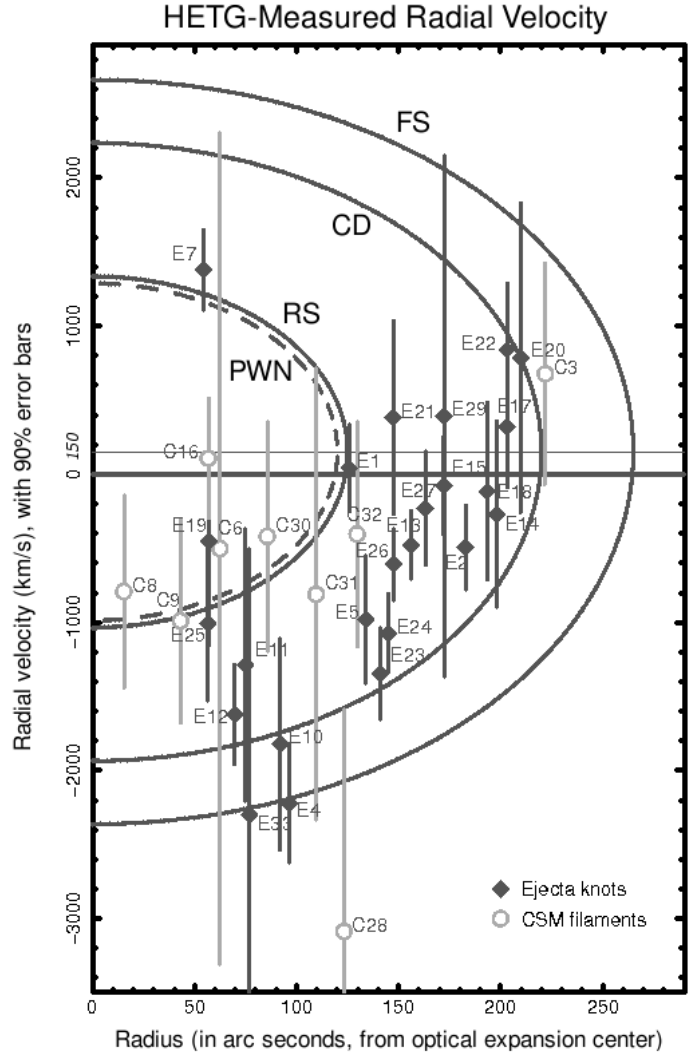}\\
\end{center}

\figcaption[]{{Radial velocity  vs.\ projected distance from the optical expansion center for 33 knots. The error bars indicate the 90\% confidence range. The elliptical loci represent expanding spherical shells of ejecta at different radial distances from the center in $v_r$\textendash $r_p$ space. Each curve is the locus of the same distance from the SNR expansion center, but with different $v_r$ observed along the line-of-sight. Radius and velocity on these elliptical loci are related through a proportionality constant based on a homologous expansion age of 3000 yr. The line at +150 km s$^{-1}$ is our estimate of the $v_r$ centroid for the elliptical loci. The dashed line roughly shows the outermost boundary of the radio PWN. The next two solid lines mark the inferred locations of the RS and CD. The outermost line at 265$\arcsec$ marks the FS.}\label{fig:Figure 3}}
\end{figure*}

\begin{figure*}[]
\figurenum{4}
\centerline{\includegraphics[angle=0,scale=0.65]{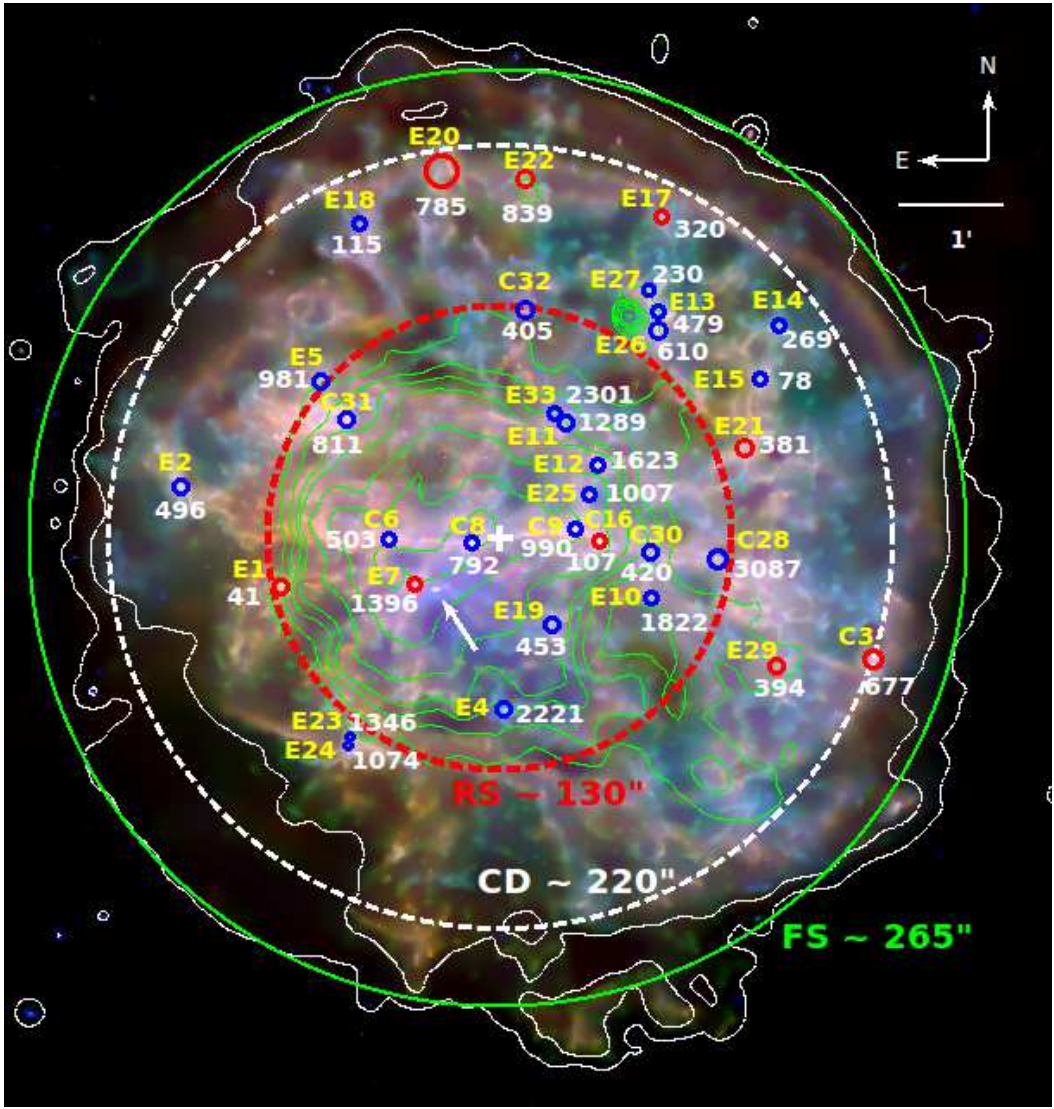}}
\figcaption[]{{ ACIS-I three-color image showing locations, identification numbers (yellow) and radial velocities (white) of 33 knots. Color codes for the image are: red = 0.3-0.8 keV, green = 0.8-1.7 keV and blue = 1.7-8.0 keV. The prefixes for the IDs are E for metal-rich ejecta, C for shocked CSM. Regions for blueshifted knots are marked with blue circles, while those for redshifed knots are marked with red circles. The optical expansion center is marked by a white cross and the pulsar PSR J1124-5916 by a white arrow. The dashed red and white circles show the locations of the RS and CD, respectively, that we infer from the ejecta distribution (Figure 3). The large green circle shows the location of the FS at 7.7 d$_6$ ($\sim$265$\arcsec$). The RS, CD and FS circles are all centered at the optical expansion center. The 20 cm map of the radio PWN is overlaid with green contours. The overlaid white contours are the outer boundary of the SNR in X-rays (based on the 0.3-8 keV broadband ACIS image).}\label{fig:Figure 4}}
\end{figure*}

\begin{figure*}[]
\figurenum{5}
\begin{center}
\includegraphics[angle=0,scale=0.65]{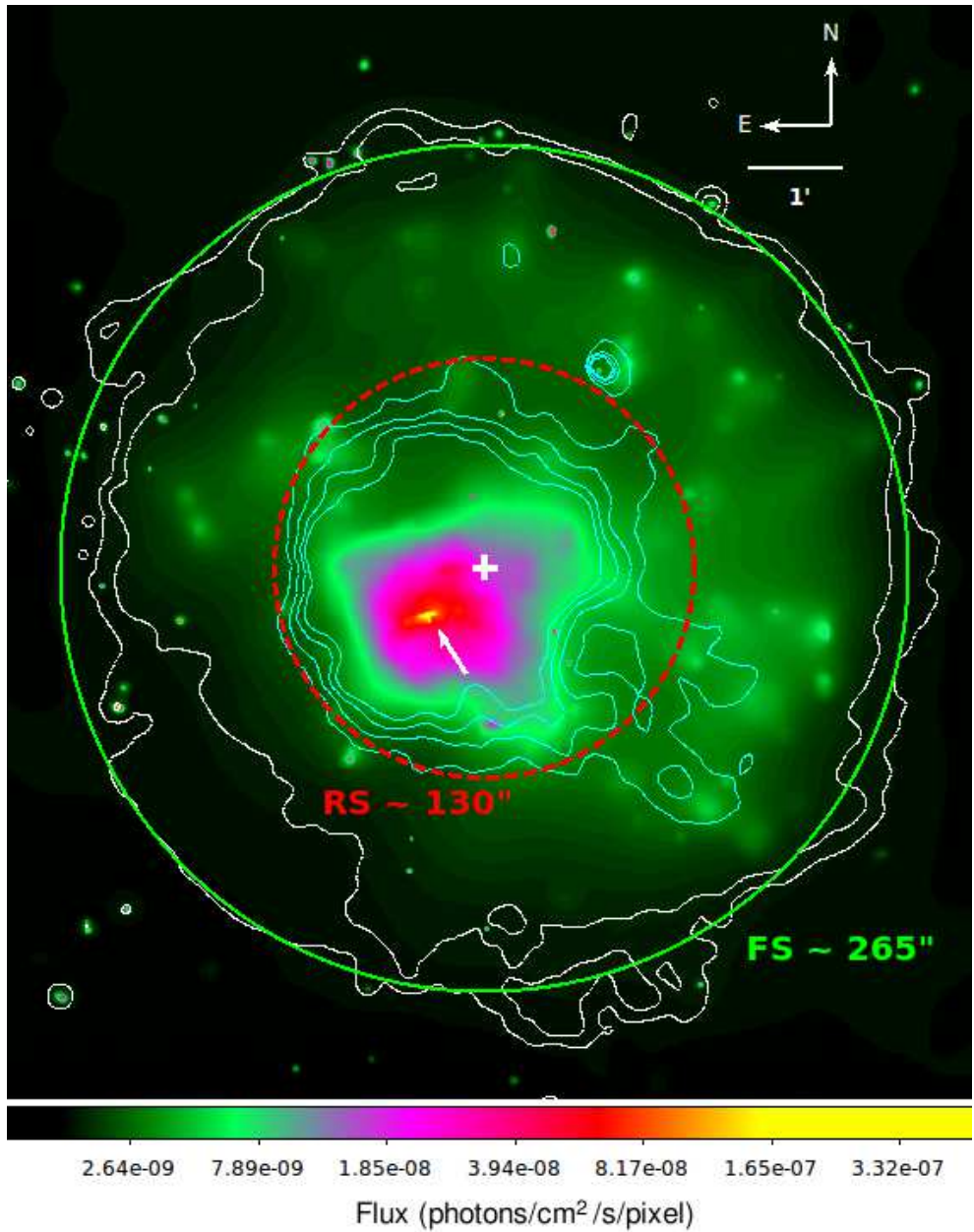}
\end{center}

\figcaption[]{{The 4-8 keV band ACIS-I image of G292.0+1.8. The image has been exposure corrected, binned by 2 $\times$ 2 pixels and adaptively smoothed. The arrow marks the pulsar PSR J1124-5916, and the cross indicates the optical expansion center. The inferred location of the RS (as marked in Figure 4) is represented by the dashed red circle at $\sim$130$\arcsec$ from the SNR's center, and the FS is indicated by a green circle at $\sim$265$\arcsec$. Overlaid are the outer contours of the 20 cm map of the radio PWN (cyan), and the X-ray contours (based on the 0.3-8 keV broadband ACIS image) marking the outer boundary of the SNR (white).}\label{fig:Figure 5}}
\end{figure*}

\end{document}